\documentclass{emulateapj}

\newcommand{\as}{\mbox{\arcsec}}

\def\lsim {$\rlap{\raise.4ex\hbox{$<$}}\lower.55ex\hbox{$\sim$}\,$}

\def\c17o{$\rm C^{17}O$}
\def\dc18o{$\rm C^{18}O$}

\begin{document}

\title {\bf Comparisons of an Evolutionary Chemical Model with Other Models} 
\author {Jeong-Eun Lee}
\affil{\it Department of Astronomy, The University of Texas at Austin,
1 University Station C1400, Austin, Texas 78712--0259}
\email{jelee@astro.as.utexas.edu}
\author {Neal J. Evans II}
\affil{\it Department of Astronomy, The University of Texas at Austin,
1 University Station C1400, Austin, Texas 78712--0259}
\email{nje@astro.as.utexas.edu}
\and
\author {Edwin A. Bergin}
\affil{\it Department of Astronomy, The University of Michigan, 825 Dennison
Building Ann Arbor, Michigan 48109-1090}
\email{ebergin@umich.edu}

\begin{abstract}
We compare an evolutionary chemical model with simple empirical models 
of the abundance and with static chemical models. We focus on
the prediction of molecular line profiles that are 
commonly observed in low mass star forming cores. We show that
empirical models can be used to constrain evaporation radii and infall
radii using lines of some species. Species with more complex abundance
profiles are not well represented by the empirical models. Static
chemical models produce abundance profiles different from those obtained from 
an evolutionary calculation because static models do not 
account for the flow of matter inward from the outer regions.
The resulting profiles
of lines used to probe infall may differ substantially.

\end{abstract}

\keywords{ISM: astrochemistry -- ISM: molecules -- stars: formation}

\section{INTRODUCTION}
Most theoretical studies of the process of star formation  
for low mass stars (Larson 1969; Penston 1969; Shu 1977; Terebey, 
Shu, \& Cassen 1984; Shu, Adams, \& Lizano 1987; Foster \& Chevalier 1993; 
Ciolek \& Mouschovias 1994; McLaughlin \& Pudritz 1997) focus on the 
formation of single stars, because these stars are simple test sites with
less observational confusion.
Such theories provide different predictions about density and velocity 
structures as a function of time.
In principle, observations in dust continuum emission and molecular transitions 
can be converted to these same physical quantities to test theory.
However, in practice the conversion is not trivial, and one should
use detailed models in a self-consistent way to correctly assess physical
conditions in star forming cores.
The density and temperature profiles of a core can be determined from
comparison of the dust continuum observations with a continuum radiative
transfer calculation. However, the velocity structure, which is the key to
the dynamics of the entire star formation process, is not easily
inferred directly from the observations of molecular transitions because
it is coupled with abundance profiles, which vary along the line of sight
due to the interactions between gas and dust grains (freeze-out and desorption
of molecules on and off grain surfaces).
Therefore, the dynamics is entangled with the chemistry.

In order to test various theoretical infall models, an evolutionary sequence of 
line profiles should be generated self-consistently to compare with a wide 
range of observations.
Lee et al. (2004, hereafter Paper I) established a self-consistent model, where
a dynamical model (the inside-out collapse model 
preceded by evolution through a sequence of the Bonnor-Ebert 
spheres) was coupled with chemistry and thermal balance in order to simulate 
the evolution of molecular line profiles.
The model has been tested by comparison with observations (Young et al. 2004; 
Evans et al. 2005). 
Alternatively, simple empirical models of step or drop functional 
abundance structures (Sch$\ddot{o}$ier et al. 2002; Lee et al. 2003; 
J{\o}rgensen et al. 
2004) have been used to fit observations.
However, according to Paper I, the abundance structures of some molecules 
are very complicated, so they are not reasonably approximated with a simple 
function.
Static models (Doty et al. 2002; Doty et al. 2004), which involve the actual
calculation of the chemical evolution but adopt constant physical conditions
with time, have been used to compare the abundance profiles across the 
envelopes with observational data and empirical profiles. 
In the prestellar stage, the static model could work because the dynamical 
timescale is long enough compared to the chemical timescale.
However, after collapse begins, the timescales for depletion and gas-phase
chemistry are comparable to the dynamical timescale (see Table 4 in Paper I). 
In addition, the evaporation of molecules modifies the gas-phase chemistry
dramatically. 
While evaporation was included in static models, they did not include the effect
of fresh material being brought into the evaporation zone.
As a result, the static models might give misleading abundance profiles.

In this paper we compare the evolutionary model established in Paper I with 
empirical models such as drop and step functions that have been
used in the literature (Lee et al. 2003; J{\o}rgensen et al. 2004;
Young et al. 2004; Evans et al. 2005). Then we compare the evolutionary
model with a static chemical model.
Our aim is two-fold: first, to explore what can be learned from the more
easily applied simple, empirical models; and second, to learn the differences
between the predictions of evolutionary models and the simpler, static models.

\section{COMPARISON WITH EMPIRICAL MODELS}
\subsection{Method}

In this section, we take the results of the evolutionary model as 
representative of the true situation and produce line profiles from them.
The line profiles are calculated for typical observing parameters and a
cloud distance of 140 pc, the distance to typical nearby regions of star
formation.
Then we try to match these line profiles by adjusting the free parameters
in empirical models to get the best match to the line profiles from the
evolutionary calculation. Then we ask how well certain parameters, such as
the evaporation radius ($R_{evap}$; \S 2.2), infall radius ($r_{inf}$; \S 2.3) 
and outer radius ($r_{out}$; \S 2.3) are determined by the fit of the 
empirical models. The fit is evaluated by using the absolute deviation (AD).
Two kinds of AD values are computed. For a given line, 
the AD is computed from $\sum_i |T_R(evol;i)-T_R(emp;i)|/N$ 
where $N$ is the number of points in the line profile.
For a species, the AD is computed from the integrated intensities over the 
transitions modeled ($AD=\sum_j |I(evol;j)-I(emp;j)|/n$ where $j$ indicates
different transitions, and $n$ is the total number of transitions).  
In almost all cases, these two measures agreed on the best fit.
We use the average of the two ADs to decide the best fit in the cases that they 
do not agree. 
C$^{18}$O and CS lines are tested for this comparison since these molecular  
lines are the most frequently used to trace low mass star forming cores, and 
the collision rates for those melecules have been well studied.

The evolutionary model used for the comparison assumes that a sequence of the
Bonnor-Ebert spheres represents the evolution of a dense core in the
pre-protostellar stage, and the core evolves based on the inside-out collapse 
model in the protostellar stage.  
The time step of $t=0$ is considered as the end of the 
pre-protostellar stage and the initiation of collapse.
For comparison to empirical models, we adopt abundance profiles at the time 
step of $t=10^5$ yr, when the infall radius is 0.0227 pc (about 33$\arcsec$ at 
140 pc) with the sound speed of 0.22 km s$^{-1}$. 
The chemical network in the model considers gas-grain interactions as well 
as the gas-phase chemistry, and it assumes bare SiO$_2$ grain surfaces and an 
external extinction of 0.5 mag (refer to Paper I for details). The binding
energy of CO onto the SiO$_2$ grain surfaces is assumed to be 1181 K 
(see Table 2 of Paper I).  

Of course, other physical, dynamical, and chemical models may describe
the actual situation, but comparison to the model described here should
provide some insights into the more general case.

\subsection{Drop and Step Models}

In prestellar cores, molecules are frozen-out onto
grain surfaces in the inner regions, so the abundance profile can be
approximated with a step function (Lee et al. 2003) with a low abundance at
small radii and a high abundance at large radii (3 free parameters per species).
This step function corresponds to the jump or anti-jump model in the 
notation of J{\o}rgensen et al. (2004).
However, once a central protostar forms, the inner regions are heated.
As a result, molecules evaporate in these regions, so abundance profiles are
assumed to become similar to a drop function. 
A drop function has three zones to describe
1) the innermost zone with a high abundance caused by evaporation of a molecule, resulting from the 
heating by the newly formed protostar 2) an intermediate zone with low (drop) 
abundance caused by freeze-out of the molecule onto grain surfaces in regions 
with low temperatures and high densities, and 3) the outermost zone with high abundance at 
large radii (the same abundance as in the inner zone), 
where densities are low, so the freeze-out timescale is longer
than the dynamical timescale.  The drop function thus has 4 free parameters
per species (the high abundance, the drop abundance,
and the inner and outer radii of the drop section). 
J{\o}rgensen et al. (2004, 2005) used drop functions 
to describe the evolution of abundance profiles in protostellar cores.
In their description, the region of evaporation becomes larger, and the drop
becomes narrower as the central protostar evolves.

Fig. 1 compares the CS 2$-$1 lines from the evolutionary model at the time steps
of $t=0$ and $t=10^5$ yr with various resolutions. 
As described above and in Fig. 1a, CS is frozen-out at small radii at $t=0$ yr, 
but at $t=10^5$ yr, CS has high abundances at small radii due to heating
by the central protostar and accretion on to the protostar.  
The evolutionary stages are more easily distinguished at higher resolution. 

The question addressed in this section is the following: how well can
the empirical models determine the evaporation radius ($R_{evap}$)?
For this test,
we adopt abundance profiles at the time step of $t=10^5$ yr from the
evolutionary model as a standard for comparison. At this time, the CO 
evaporation radius in the model, $R_{evap} = 0.003$ pc, corresponding to 
4\farcs4 at 140 pc.
We have then tuned drop models 
to fit line profiles simulated with the standard evolutionary model 
as well as possible.  
We did not explore a complete grid of models. Instead, we first chose 
the four parameters of the drop function based on the evolutionary model
and tuned the parameters to find a reasonably good fit.  
This process simulates a procedure in which a model of continuum emission
provides a dust temperature distribution and knowledge of evaporation
temperatures is used to select the inner boundary of the drop region.
Once the first good fit model was obtained, we varied each parameter by some 
factors to check that the fit was still the best, or to find a better fit.  

The CO abundance profiles of the two models are plotted in Fig. 2a. 
In the figure, at large radii, the high abundance of the drop function is 
similar to the normal abundance of the evolutionary model. 
In addition, the best-fit inner boundary of the drop zone 
($R_{evap}$) is 
located almost at the same radius as the evaporation front in the standard 
evolutionary model. 
Fig. 2c compares the line profiles of three C$^{18}$O transitions simulated 
with the abundance profiles in Fig. 2a with the resolutions that have been used 
in J{\o}rgensen et al. (2002).
The line profiles from the two models are almost identical.
According to the tests of various drop functions, we found that the
CO evaporation front, which is located at 0.003 pc, could be well constrained
by the simple model (Fig. 2b).
If we use evaporation radii smaller and larger than that (0.003 pc) of the 
best model by a factor of two, that is, 0.0015 and 0.006 pc, the absolute 
deviations calculated by comparisons 
between line profiles simulated from the standard model and drop models with 
the two evaporation radii are greater than that of the best drop model by 
factors of about 5 and 7, respectively.  
The biggest variation of AD occurs in the 3$-$2 line because the higher 
C$^{18}$O transition has the highest critical density, and the resolution for 
the 3$-$2 line is smallest, so the 3$-$2 line is the most sensitive to the 
size of the high abundance zone at small radii.
As a result, the CO evaporation front, which determines the size of
the high abundance zone in the regions with high densities, can be well
constrained by drop models if the proper transition and adequate resolution 
are used. 

The CS 2$-$1 lines with various resolutions are shown in Fig. 3.
In this comparison, we have tested both step and drop functions, and Fig. 3a
shows the drop and step abundance profiles that best match the evolutionary
model at $t=10^5$ yr.
The results from the drop and step models shown in Fig. 3a are compared to
the result from the evolutionary model in Fig. 3b and 3c, respectively.
Drop functions cause self-absorptions to be deeper than the evolutionary
model because of high abundances at large radii.
Step functions work better for the self-absorption dips in the CS 2$-$1 line in
this comparison since the less attenuated ISRF reduces the CS abundance at 
large radii significantly.
However, as seen in Fig. 4 (Fig. 13 of Paper I), the CS abundance structure 
resulting from the model with 3 mag of external extinction is more closely 
approximated with a drop function because it has higher abundance at the outer
radii, caused by lower photodissociation by interstellar UV photons.  
If higher spatial resolution is used, neither a step function nor a 
drop function fits the high velocity wings (the third panels of Fig. 3b and 3c)
that are produced by the high abundance at very small radii. 
Of course, the high velocity wings from
infall may be confused by outflow in a real source.

There are two jumps of the CS abundance profile in the evolutionary model; 
one is associated with CO evaporation at 0.003 pc, and the other is 
caused by the evaporation of CS itself at 0.001 pc.  
In both functions, the CO evaporation front is well constrained by the CS
line as it was in the CO line profiles. 
The better constraint is given by the higher resolution observation.
The critical density of CS 2$-$1 is about $2\times 10^5$ cm$^{-3}$, and
the density around the CO evaporation radius at $t=10^5$ yr is not very 
different from the critical density, so the 2$-$1 line, even with lower 
resolution, is affected by the location of the CO evaporation front.
The CS evaporation front 
cannot be constrained with this kind of simple step or drop function. 
The evaporation of CS from grain surfaces occurs at very small radii ($r< 
0.001$ pc) with high temperatures, so if the abundance jump of CS 
at the CO evaporation radius is ignored, 
the intensity and width of simulated lines are too 
small compared to the evolutionary model.  Functions with two steps and 
observations with high resolution might be able to constrain the 
CS evaporation front as well as the CO evaporation front.

The difference between the evolutionary model and the empirical model is more 
significant in transitions from higher energy levels at high spatial 
resolutions.
Fig. 5 shows the comparison of the CS 5$-$4 lines from the evolutionary
and step function abundance profiles. 
The red peak and blue wing around $\pm 1.5$ km s$^{-1}$ are caused by the
abundance peak at $R_{evap}$(CS) ($\leq 0.001$ pc).
The abundance peak also increases the peak radiation temperature by a few K.
Even though the volume of the inner small region where CS has its abundance
peak is very small, high density and temperature combined with the abundance
peak cause high excitation of CS.

\subsection{Variations in the Physical Model}

In this section, we address the question: how well can the empirical
models determine the infall radius ($r_{inf}$) and the outer radius
($r_{out}$)? For this test, we retain the same evolutionary model
at $t = 10^5$ yr, which produces an infall radius, $r_{inf} =  0.0227$ pc,
with an outer radius, $r_{out} = 0.15$ pc,
but we change the relevant radii in the empirical model by a factor of
two in either direction to see if the fit to the observations degrades. 

First we change the infall radius and simulate the CS 2$-$1 line.
In this test, we assumed the same luminosity (about 1 L$_\odot$), which can be
constrained by observations, and kinetic temperature structures were
calculated self-consistently.
The CS 2$-$1 line has been simulated with a resolution of 30$\arcsec$.
In Fig. 6, profiles drawn with the dotted line show the best step models with
different infall radii compared to the evolutionary model at $t=10^5$ yr.
The physical models with different infall radii are illustrated in Fig. 7.
In Fig. 7d, abundance profiles of the three best step models are compared with
that of the evolutionary model.
We tested different higher and lower abundances in step models, but all the best
step models finally had the same higher ($4.0\times 10^{-9}$) and lower
($4.0\times 10^{-10}$) abundances.
The inner region of the CS higher abundance is smaller in the model with
the smaller infall radius, and vice versa.
The higher and lower densities in the model with the smaller and larger infall
radii produce stronger and weaker CS 2$-$1 lines, respectively, as shown
in Fig. 6.
The model with the smaller infall radius has smaller velocities (Fig. 7c) within
the infall radius, so the simulated line profile is narrower than the
CS 2$-$1 line profile from the evolutionary model (the first panel of Fig. 6).
On the other hand, the model with the larger infall radius produces broader
line profiles (the third panel of Fig. 6).
These trends are also found in the higher transitions of CS, which can trace 
denser regions, simulated with the spatial resolutions that are currently 
available.  
Based on this test, we think that the infall radius of a core collapsing 
from the inside-out might be constrained by a simple model such as the step 
model with optimized abundances. 
This result is consistent with that of Evans et al. (2005), who
found that evolutionary and step function models agreed on the $r_{inf}$
needed to match observations of CS in B335.

The determination of the outer radius has been also tested with CS 2$-$1.
We decreased and increased the outer radius by a factor of 2 in the step model.
In our standard evolutionary model, the outer radius is 0.15 pc.
The infall radius is the same (0.0227 pc) in all models, so the velocity
structure is also the same in all models.
The kinetic temperature structure for each model has been calculated
self-consistently, and the abundance profile in three different models are
the same step function as shown in Fig. 3a.
Fig. 8 compares the line profiles simulated from models of different outer
radii with the same luminosity, infall radius, and step abundance profile
with the line profile from the evolutionary model at $t=10^5$ yr.
The result does not show big differences in models with different outer radii.
We also used C$^{18}$O transitions, which trace less dense regions better than
CS transitions, for this test and found that C$^{18}$O might constrain only
the minimum value of $r_{out}$ with the $1-0$ line.
However, the higher transitions of C$^{18}$O, as with CS lines, did not show 
big differences in models with different outer radii. 
We tested the line ratios of CS and C$^{18}$O as a constraint of the outer 
radius.
The line ratios of CS were similar for different outer radii, but the line
ratios of C$^{18}$O showed the possibility of constraining the outer radius.
If we use outer radii smaller and larger than 0.15 pc by a factor
of two, that is, 0.075 and 0.30 pc, the ADs are greater than that of the 
model with $r_{out}=0.15$ pc by factors of 4.5 and 2.5, respectively. 
Therefore, in order to constrain the outer radius of an isolated core, 
one might have to observe multiple lines of appropriate molecules such as 
C$^{18}$O and calculate their line ratios. 
In the case of star forming cores associated with bigger molecular clouds,
however, C$^{18}$O is not a good choice because the $1-0$ line emission will be
contributed partly by the extended molecular cloud, so even its line ratios 
will not constrain outer radii.

We also tested the step model to constrain the infall radius and outer
radius with CS lines at a different time step, $t=3\times 10^5$ yr. 
The infall radius at that time step is about 0.0703 pc. 
Using this test, we find the same result as before at $t=10^5$ yr, that the 
outer radius is not well constrained by the simple model of CS. 
However, we could constrain only the minimum infall radius with the step model 
of CS at $t=3\times 10^5$, unlike the result at $t=10^5$ yr.  
At later times, the infall radius is large, and the densities around 
the infall radius are too low to be traced well by the CS lines. 
Therefore, other molecular lines that have lower critical densities need to be
used to constrain the infall radius in later evolutionary stages.
   
\section{COMPARISON WITH STATIC MODELS}

The next step beyond empirical models is a time dependent chemical model
calculated for a static physical model.
Doty et al. (2004) used density and temperature structures obtained by 
Sch$\ddot{o}$ier et al. (2002) in order to test the chemical distribution 
across the envelope of IRAS 16293-2422.
Even though they calculated  the chemical evolution, the physical (density and
temperature) structures were not varied with time. 
As a result, the evaporation fronts of molecules, which affect molecular line 
profiles significantly, stay at the same radii with time.
However, the density structure should change with time in any dynamical theory,
and the temperature structure varies with time depending on the 
accretion rate and mass of the star (thus the evolution of luminosity).
In Paper I, we have adopted Shu's picture as the standard for the dynamical 
evolution providing a constant accretion rate, so the central luminosity, thus
the dust temperature, increases ($L\propto M_\ast$). 
As a result of the evolution of the temperature
structure, the evaporation fronts propagate outward with time.  
In this study, therefore, we compare the evolutionary model with two flavors of 
static models: a truly static model and a partly static model.
The truly static model does not have the variation in physical conditions as in
Doty et al. (2004).
In the partly static model, we consider the evolution of physical conditions 
based on Shu's inside-out collapse model and the results of dust continuum 
radiative transfer calculations without following each gas parcel.

Fig. 9 shows the comparison between the evolutionary model and the truly static
model.
We adopt density and temperature structures from the evolutionary model at the
time step of $t=10^5$ yr, as shown in Fig. 7a and 7b with the solid lines. 
For this comparison, we use the standard chemical model in Paper I, 
where the molecular binding energies onto the bare SiO$_2$ grain 
surfaces are considered and the ISRF is attenuated by 0.5 mag. 
The same initial abundances, shown in Table 3 of Paper I, have been used in 
both the evolutionary model and the truly static model.
The left and right panels of Fig. 9 compare the abundance profiles from the 
truly static model at $t=10^5$ and $t=1.1\times 10^6$ yr, respectively, 
with those of the evolutionary model at the time step of $t=10^5$ yr. 
In the evolutionary model, the model core spent $1\times 10^6$ years in the
pre-protostellar stage, so the whole timescale for the chemical evolution at
$t=10^5$ yr is $1.1\times 10^6$ years.
CO, CS, H$_2$CO, and HCO$^+$ are molecules that reach their maximum abundances
fast, so they are already in the phase of depletion at $t=10^5$ yr.
However, nitrogen-bearing molecules such as NH$_3$ and N$_2$H$^+$ are
related to N$_2$, which forms slowly, so those molecules become more abundant 
even until $t=1.1\times 10^6$ yr.  
In the truly static model, the abundances of those nitrogen-bearing molecules
never reach the maximum values within the CO evaporation radius because the 
abundant CO will destroy them directly or indirectly.
The biggest difference between the evolutionary model and the truly static 
model in the right panel is shown in the CS abundance profiles; a deep 
depletion zone between $R_{evap}(\rm CS)$ and $R_{evap}(\rm CO)$ in the static 
model. 
This difference is due to the dynamical effect.
In the evolutionary model, the material with a high CS abundance in the outer
region around $R_{evap}(\rm CO)$ continues to flow into the zone to cancel out 
the depletion effect, but in the static model, the depletion of CS occurs 
continuously from the same material to reduce the abundance more with time in 
that zone.
The other difference is the occurrence of peaks inside the CO evaporation 
radius.
For example, for H$_2$CO and HCN, the peaks that appear in the evolutionary 
model disappear in the truly static model because no fresh material is brought
in to radii smaller than $R_{evap}(\rm CO)$ to sustain the peaks in the 
static model, unlike in the evolutionary model.  

In the partly static model, we simply assume that the physical conditions evolve
without any dynamical process, as in Bonnor-Ebert spheres, and calculate
the chemical evolution at given radii in an Eulerian coordinate.
In contrast, in the evolutionary model of Paper I, the chemical evolution has 
been calculated in a Lagrangian coordinate system following all gas parcels as 
they fall toward the center and transferred to an Eulerian coordinate system to 
provide abundance structures at any time step. 
Even though we see the same physical conditions at a given radius at a given 
time step in the two models, the histories of physical conditions that a gas
parcel in the two models goes through until the given time step are totally 
different. 
As illustrated in Fig. 10, in the inside-out collapse, the density evolves from
lower to higher while the chemical evolution of a given gas parcel is 
calculated in the evolutionary model. On the contrary, in the partly static 
model, the chemical evolution is calculated at a given radius, so the density 
evolves from higher to lower.  
In addition, the evolutionary model always experiences a lower temperature than 
the partly static model does until the given time step after the central 
protostar forms, since the gas parcel in 
the evolutionary model is located at a radius greater than the radius where 
the partly static model is calculated at all times until the given time step.
Thus, the past history of physical conditions affect the depletion and 
evaporation of molecules, causing differences in their abundance structures 
and, finally, line profiles. 

For the comparison between the evolutionary model and the partly static model, 
as described above, we use the standard chemical model in Paper I. 
The partly static model spends $10^6$ years in the pre-protostellar stage 
passing through a sequence of Bonnor-Ebert spheres before collapse, as does the 
evolutionary model. Therefore, the initial abundances for the chemical 
evolution after collapse are the same in the two models.   
Fig. 11 shows the differences of abundance profiles resulting from
the evolutionary model and the partly static model at $t=10^5$ yr after collapse
begins. 
First, outside the infall radius (about 0.023 pc), the two models show 
identical abundance profiles since no dynamical motion is involved. 
However, the partly static model shows more freeze-out
of molecules between the infall radius and the CO evaporation front, due to
the history of higher densities in the static model.
In the evolutionary model, a gas parcel 
moves from an outer radius to the given radius carrying less frozen-out 
material.
In the static model, the evaporation fronts are also located at slightly 
larger radii than in the evolutionary model because the static model always 
experiences higher temperatures than does the evolutionary model after the 
central protostar forms, and the time interval between two time steps is not 
infinitesimal in the evolutionary calculations.
The biggest difference between two models lies  in the abundance peaks caused 
by the evaporation of molecules other than CO.
In the static model, the peaks are very weak or disappear.
The difference is caused by the dynamical effect; that is, in the static model,
once molecules evaporate and reach chemical
equilibrium, no more evaporation occurs from grain surfaces.
However, in the evolutionary model, material is being fed into these regions to
sustain high abundance peaks.
This material effectively produces an abundance pile-up at the evaporation front
for a given volatile as the grains sweep past.
The structures in the evolutionary model are also smeared in radius as a result 
of the dynamical effect. 
These differences will be more significant in other dynamical models with
higher velocities.
We also tested the case of water dominant grain surfaces, and 
the results show that the difference between the static model and the 
evolutionary model is more significant because more freeze-out is caused 
by higher binding energies.

Among molecules in Fig. 11, the most significant difference between the partly
static model and the evolutionary model is seen in CN, 
so one can expect the greatest difference in CN line profiles too.
However, we do not have good collision rates for the transitions of CN; we 
therefore focus on H$_2$CO instead. 
In Fig. 12, we compare H$_2$CO line profiles resulting from abundance profiles
of the two models to show that the ratio between blue and red peaks, which 
increases as the infall speed increases, is smaller in the static model if a 
high resolution (3\as) is used. 
In addition, as a result of the inner zone of the high H$_2$CO abundance 
in the evolutionary model, the line profile from the evolutionary model is 
broader than that from the static model. The peak intensities are also 
stronger by about a factor of 2 in the evolutionary model.  
Therefore, it is possible to infer faster infall 
velocities than the actual infall velocity by using the static model. 
In the simulation of line profiles, we adopt the velocity structure of the
inside-out collapse at $t=10^5$ yr for both the evolutionary model and the 
static model even though the static model assumes no dynamical motion
in the chemistry.

The difference between the partly static and evolutionary models occurs within 
the infall radius, which increases with time, so the difference can only 
be caught by molecular observations with very high resolutions, 
in the earlier evolutionary stages. 
Therefore, the velocity structure is entangled with abundance structures at 
small radii, which might be possibly connected to disk dynamics as well as 
envelope dynamics, but it will be constrained by telescopes such as CARMA and 
ALMA, which will become available in the near future.  

\section{CONCLUSIONS}

Although actual abundance profiles cannot be matched 
completely by the empirical models, our comparisons suggest the empirical 
models may fit line profiles of CO and its isotopes.  
CO has a relatively simple abundance profile and hence can be more readily 
matched.  However, the
empirical models are very different for other molecules (e.g., CS, H$_2$CO, CN),
which have more complicated abundance profiles in the inner zones (Paper I).
For these more complicated models the disparity between the
empirical model and the true profile is magnified
with higher spatial resolution in higher energy level transitions.
However, we can still learn about $R_{evap}$ and $r_{inf}$ 
from the empirical models applied to appropriate lines.
The CO evaporation radius can be constrained in the empirical models
if the temperature structure is well known through dust continuum radiative
calculation. The CO evaporation temperature depends on the surface properties
of dust grains.
For example, if the grain surfaces are covered dominantly by water, then the
evaporation temperature should be higher than in the case of the CO-dominant
grain surfaces.
In addition, if the inside-out collapse model is the correct theory for  
isolated star formation, the simple models can constrain 
the infall radius in a star forming core.
$r_{out}$ might be also constrained with the empirical models if the line 
ratios of appropriate molecules are used.  

In this comparison, we should not exclude the possibility that the 
discrepancy between the empirical model and the full evolutionary model may be 
a signal that something is missing or incorrect in the chemical description of 
a specific species, such as an incorrect binding energy, missing reaction, 
and so on. 
For example, we did not include surface chemistry, which might be very 
important in the chemistry of H$_2$CO. 
Sch$\ddot{o}$ier et al. (2004) found
that a drop abundance profile fitted well the the ratios of many lines and 
the interferometer data of H$_2$CO toward L1448-C and IRAS 16293-2422. 
In contrast, Evans et al. (2005) showed that neither the evolutionary 
chemical model nor any step model matched high J lines of H$_2$CO in B335. 

We tested two static models; one is the truly static model in which 
physical conditions do not vary with time, and 
the other is the partly static model that includes the evolution of physical
conditions, but does not consider the dynamical effect in the chemistry. 
Although the partly static model is much 
closer to the evolutionary model than the truly static models that previous 
studies have used, the comparison between the evolutionary model and the partly
static model shows that the static models have different profiles than the 
evolutionary ones, with the differences becoming more apparent at high 
resolutions and for high energy transitions.
As mentioned in previous sections, the difference between the evolutionary model
and the static model will be dependent on the velocity structure of a star 
forming core. Therefore, we have to use high resolution and high-excitation
lines to constrain the velocity structure coupled with the abundance profiles. 
Such constraints are necessary to decide which theoretical model best explains
the star forming process.
These observations will be practicable with the CARMA and ALMA in the near 
future. 

\acknowledgments

We thank E. van Dishoeck for suggesting these comparisons and helpful comments. 
We also thank J. J{\o}rgensen for useful discussions.
This work was supported in part by NSF grant AST-0307250 to the University
of Texas at Austin and AST-0335207 to the University of Michigan.
Jeong-Eun Lee is grateful to the Department of Astronomy and the University
of Texas at Austin for their support through the David Allen Benfield Memorial 
Scholarship and the University of Texas at Austin Continuing Fellowship.

\clearpage

\begin{figure}
\figurenum{1}
\epsscale{.4}
\plotone{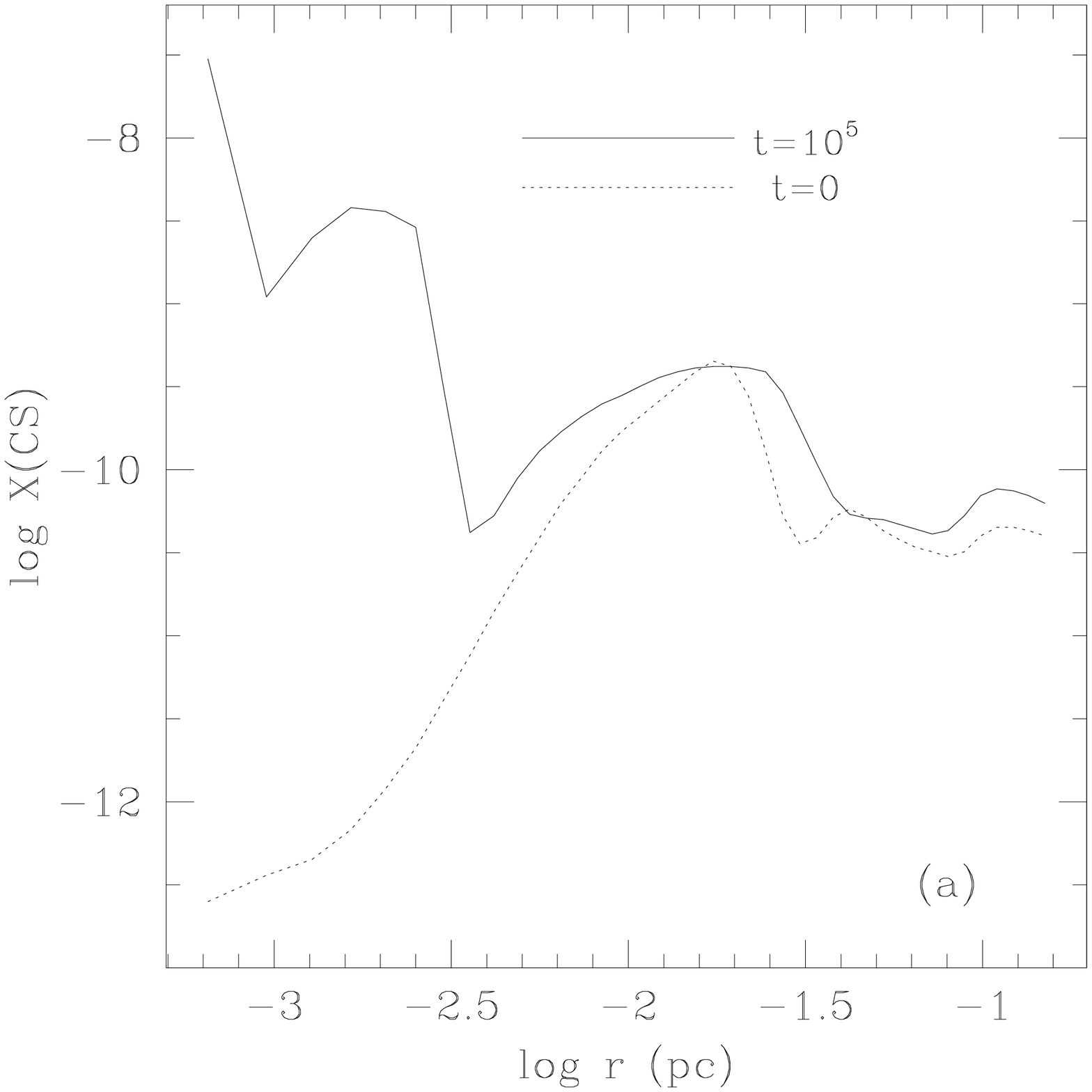}\plotone{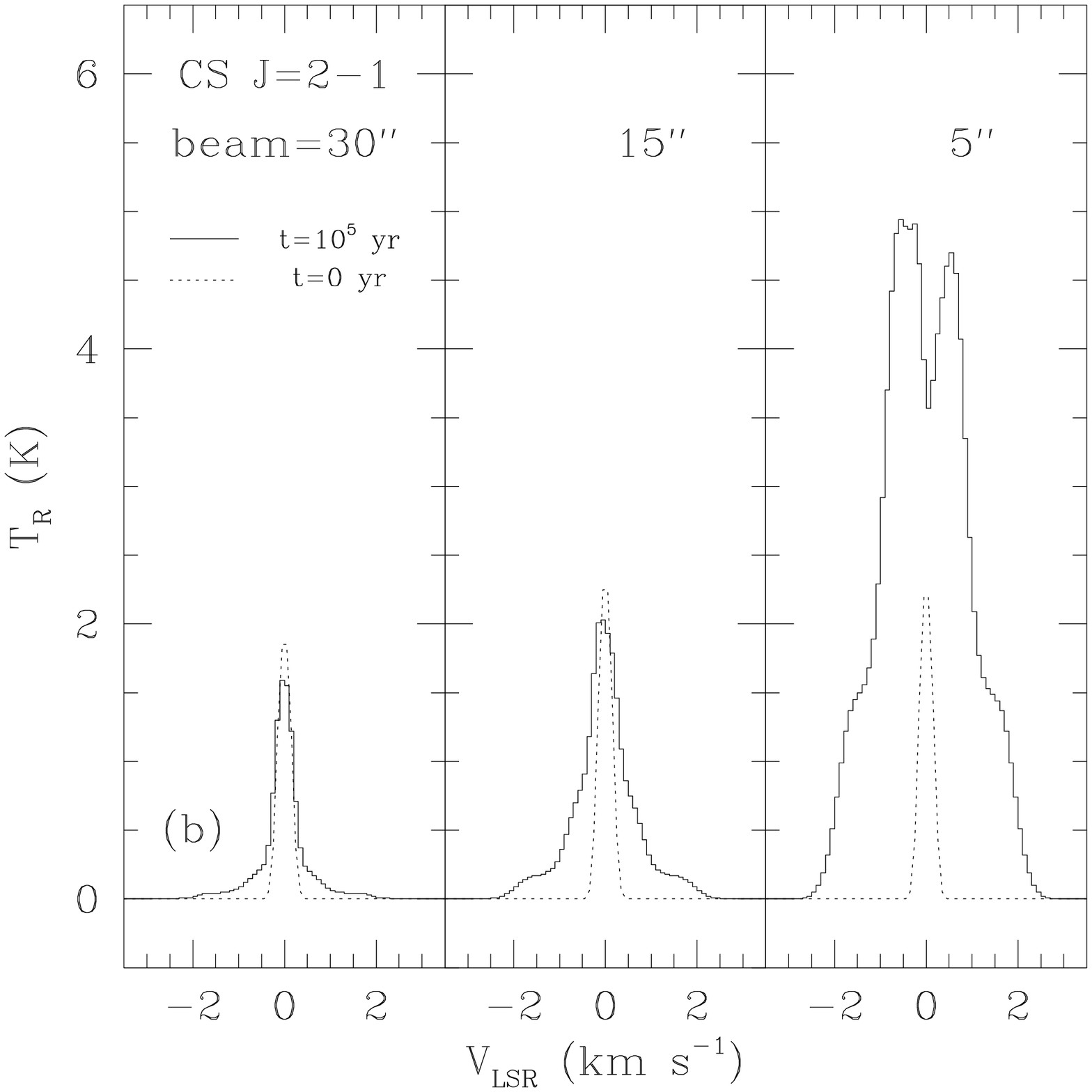}
\caption{
Comparison of the CS 2$-$1 lines from the evolutionary model at $t=0$ and 
$t=10^5$ yr; 
(a) the comparison of abundance profiles at two time steps (solid line 
for $t=10^5$ and dotted line for $t=0$)
and (b) comparison of the CS 2$-$1 line profiles calculated from the 
evolutionary model at $t=0$ (dotted line) and $t=10^5$ (solid line).
The evolutionary stages can be distinguished better at higher resolution.
}
\epsscale{1.0}
\end{figure}

\begin{figure}
\figurenum{2}
\epsscale{.4}
\plotone{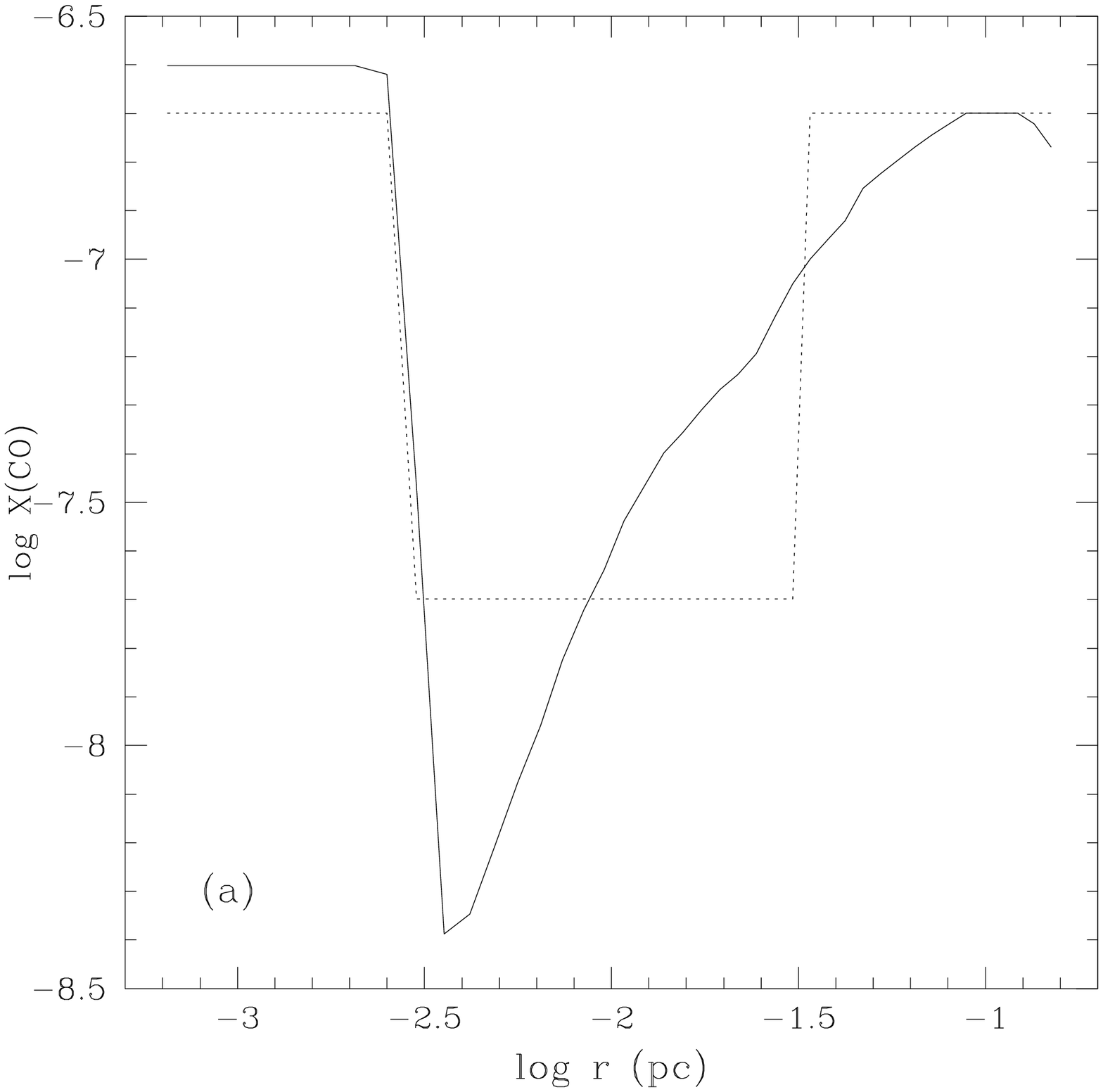}\plotone{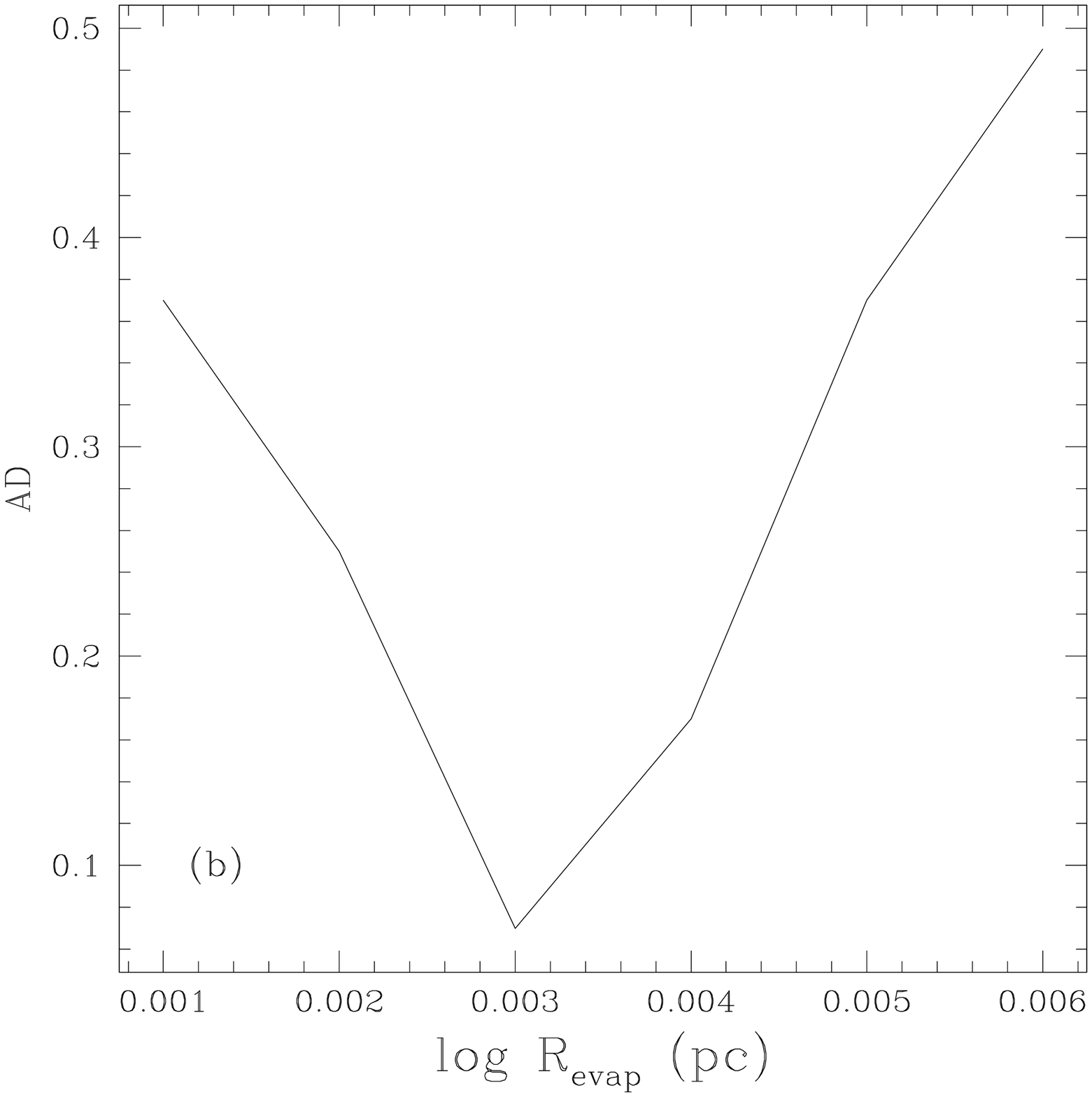}
\plotone{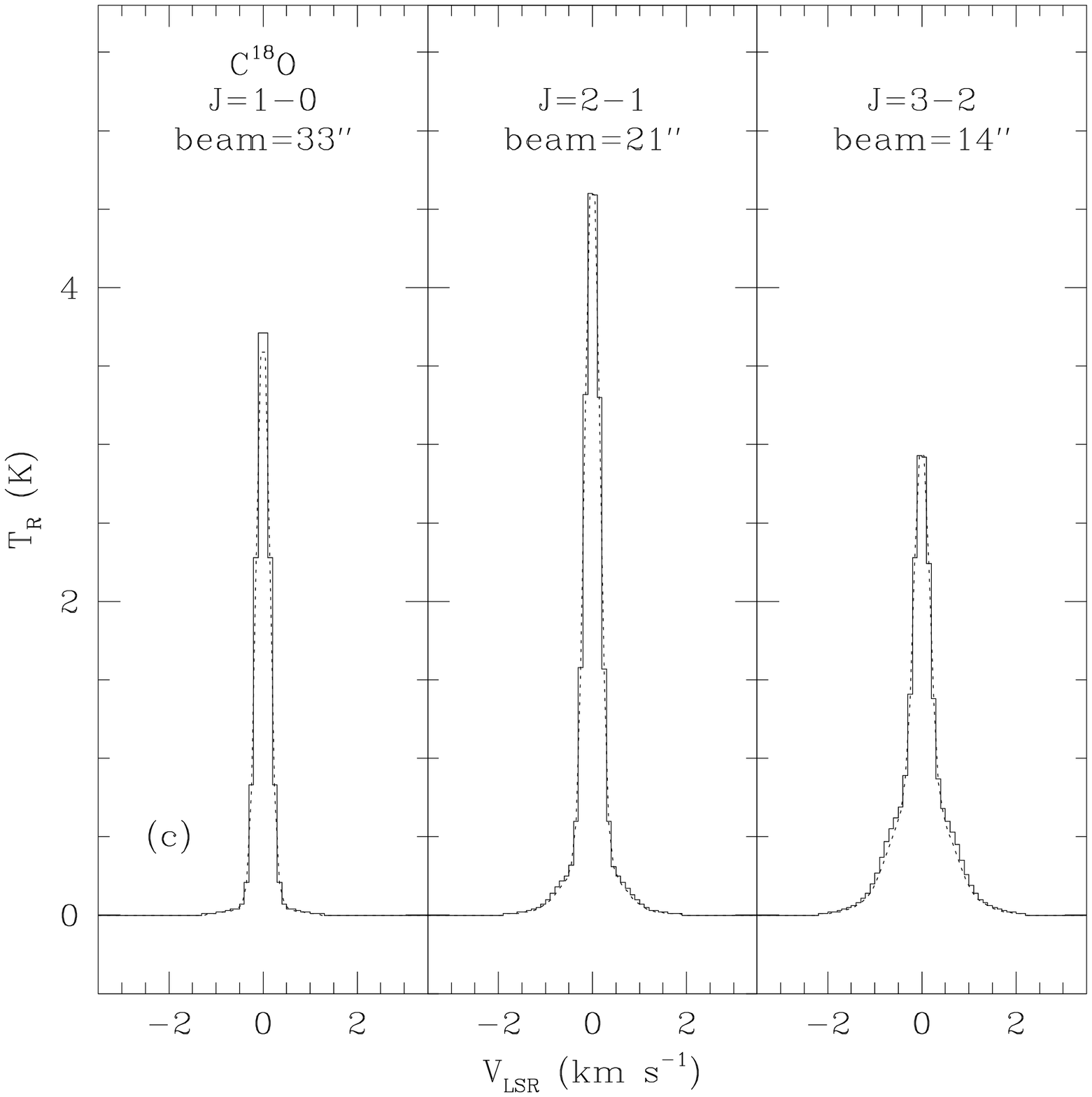}
\caption{
Comparison between the evolutionary model at $t=10^5$ yr and a drop model
in C$^{18}$O;
(a) the comparison of abundance profiles in two models (solid line for the
evolutionary model and dotted line for the best-fit drop model to the 
evolutionary model),
(b) the absolute deviation of integrated intensities vs.
the evaporation radius in the drop model, which is calculated over 
three C$^{18}$O line profiles simulated with two abundance models,
and (c) the comparison of the C$^{18}$O line profiles of the evolutionary model
(solid line) and the best-fit drop model (dotted line) to the evolutionary model
in three different transitions.   
}
\epsscale{1.0}
\end{figure}

\begin{figure}
\figurenum{3}
\epsscale{.4}
\plotone{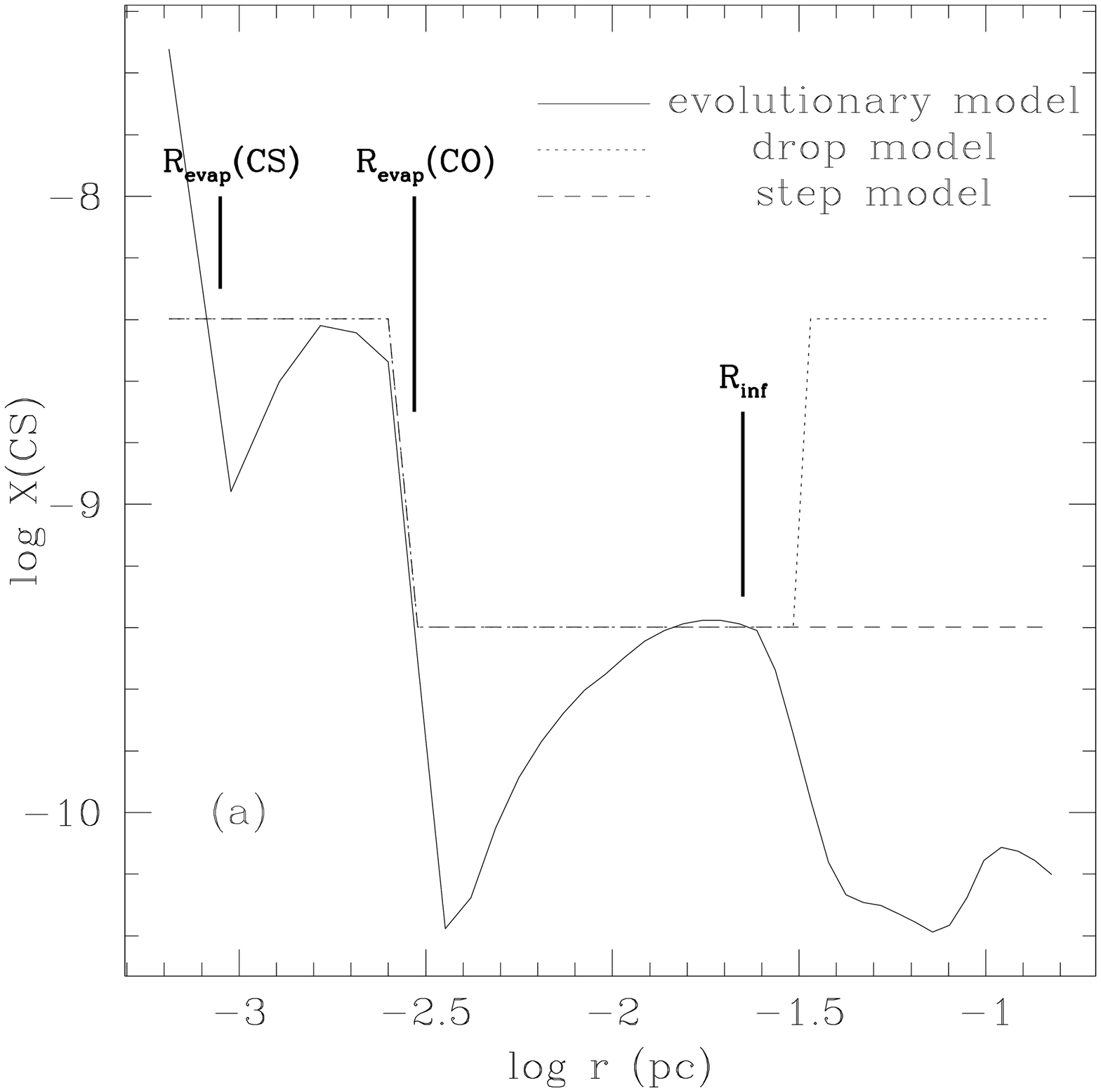}\plotone{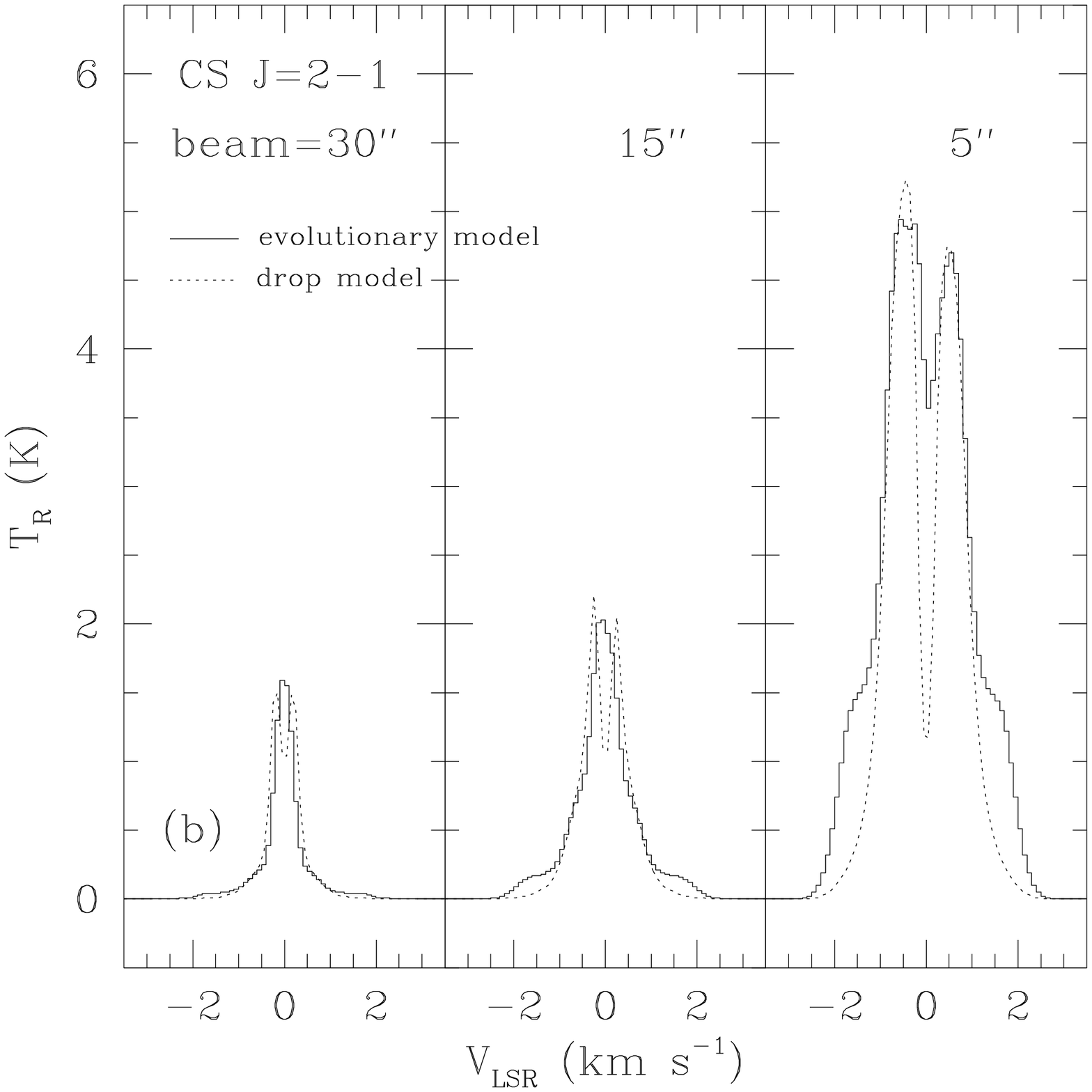}
\plotone{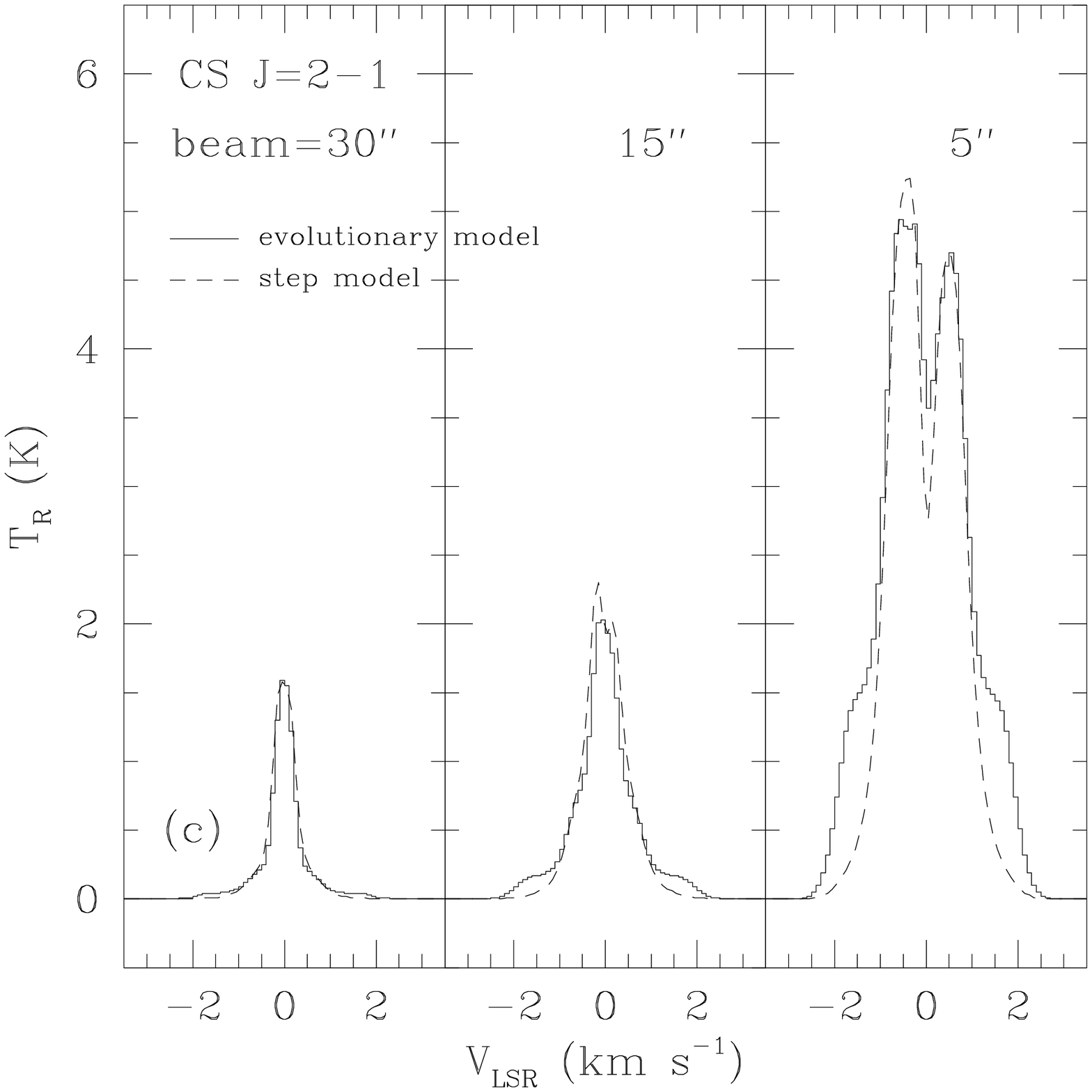}
\caption{
Comparison between the evolutionary model at $t=10^5$ yr, a drop model, and
a step model in CS;
(a) the comparison of abundance profiles in three models (solid line for the
evolutionary model and dotted and dashed lines for the best-fit drop and step
models to the evolutionary model, respectively),
(b) the comparison of the CS 2$-$1 line profiles of the evolutionary model
(solid line) and the best-fit drop model (dotted line) to the evolutionary model
with three different resolutions.
and (c) the comparison of the CS 2$-$1 line profiles of the evolutionary model
(solid line) and the best-fit step model (dashed line) to the evolutionary model
with three different resolutions.
In (a), thick vertical lines indicate the evaporation radii of CS and CO.
}
\epsscale{1.0}
\end{figure}

\begin{figure}
\figurenum{4}
\epsscale{.4}
\plotone{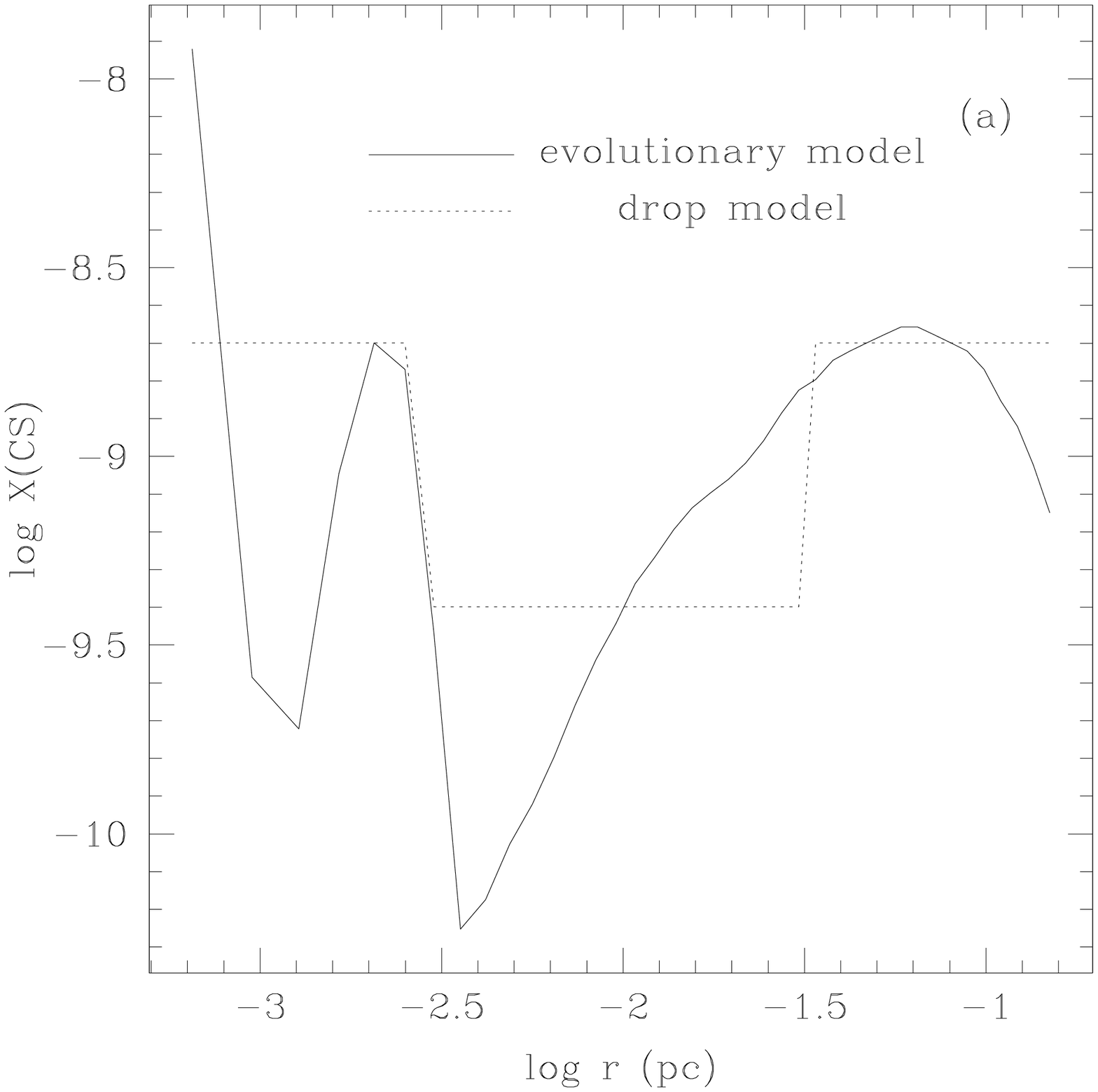}\plotone{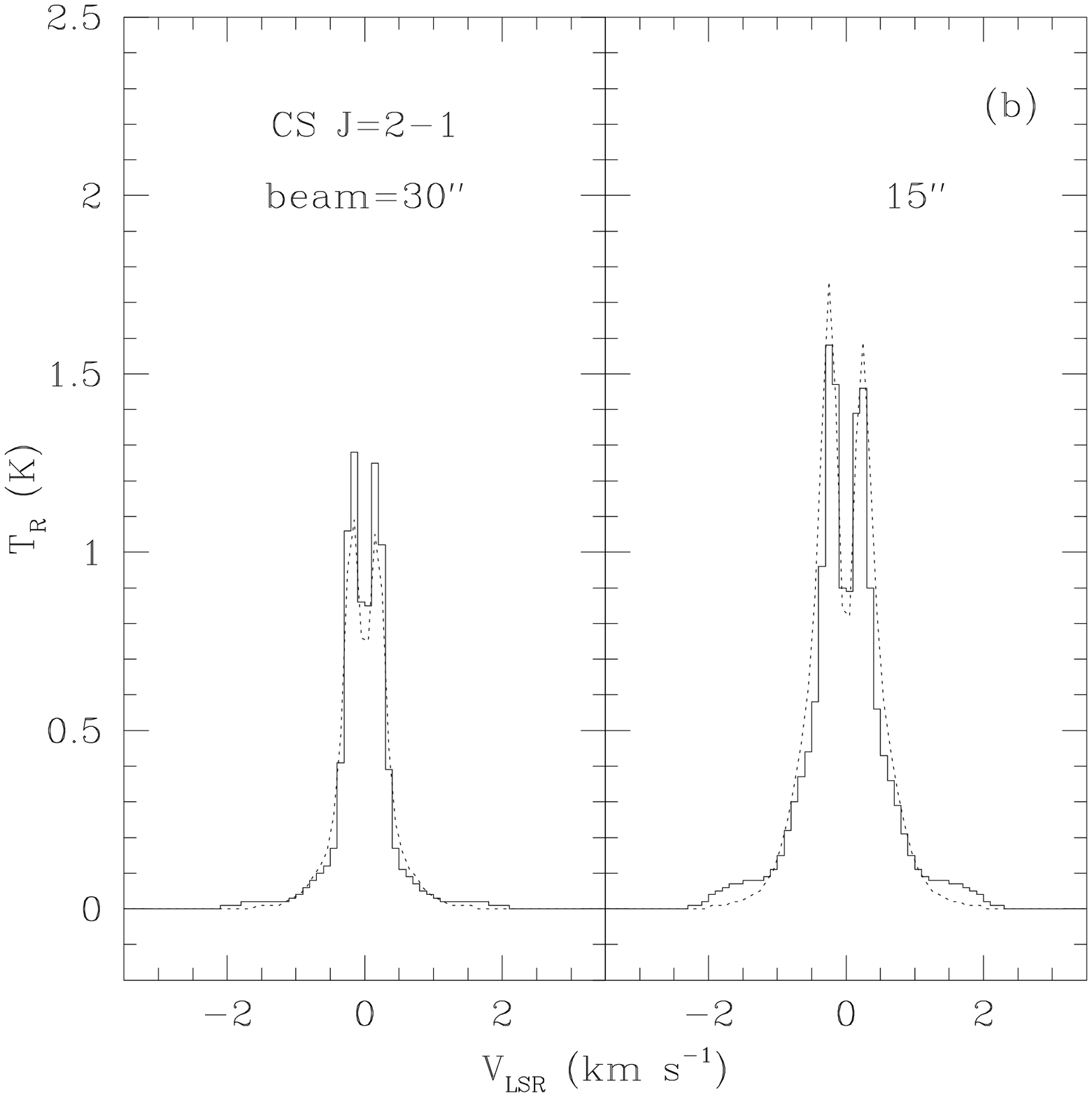}
\caption{
Comparison between the evolutionary model with an external extinction of 3 mag
at $t=10^5$ yr and a drop model in CS;
(a) the comparison of abundance profiles in two models (solid line for the
evolutionary model with 3 mag of external extinction and dotted line for 
the best-fit drop model to the evolutionary model)
and (b) comparison of the CS line profiles calculated from the evolutionary 
model with an external extinction of 3 mag (solid line) and the best-fit drop 
model (dotted line) to the evolutionary model with two different resolutions.
}
\epsscale{1.0}
\end{figure}

\begin{figure}
\figurenum{5}
\epsscale{.6}
\plotone{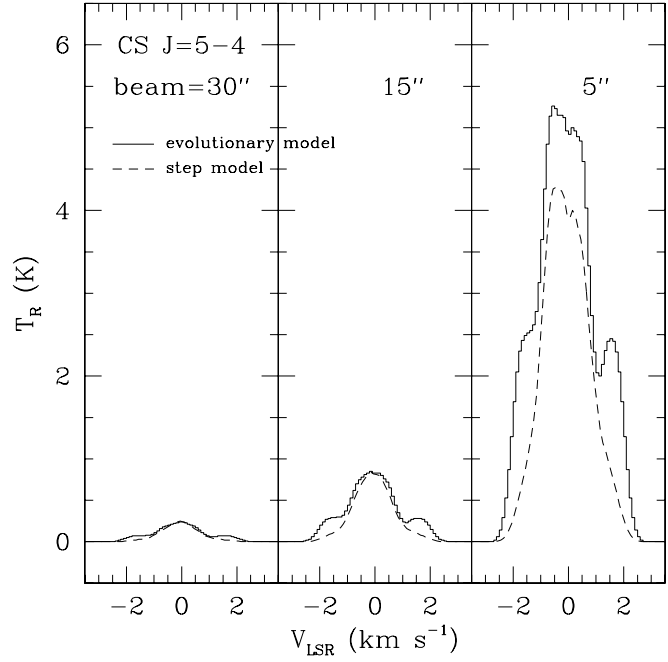}
\caption{
The comparison of the CS 5$-$4 line profiles simulated with the abundance 
profiles of the evolutionary model (solid line) and the step model (dashed line)
in Fig. 3a. 
}
\end{figure}

\begin{figure}
\figurenum{6}
\epsscale{.6}
\plotone{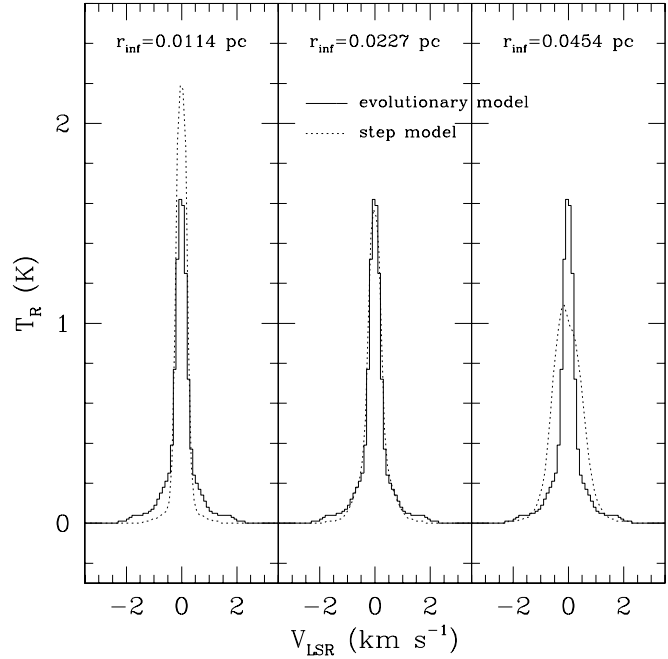}
\caption{
The CS 2$-$1 line profile (solid line) simulated from the evolutionary model
at $t=10^5$ yr is compared with profiles (dotted line) predicted by step models
with different infall radii.
The beam size for this test is 30$\arcsec$.
The first panel shows the best model with $r_{inf}=0.0114$ pc; the second panel
shows the best model with $r_{inf}=0.0227$ pc, which is the infall radius
at $t=10^5$ yr; the third panel shows the best model with $r_{inf}=0.0454$ pc.
The second panel shows the same comparison as the first panel in Fig. 3c
does.
}
\end{figure}

\begin{figure}
\figurenum{7}
\epsscale{.5}
\plotone{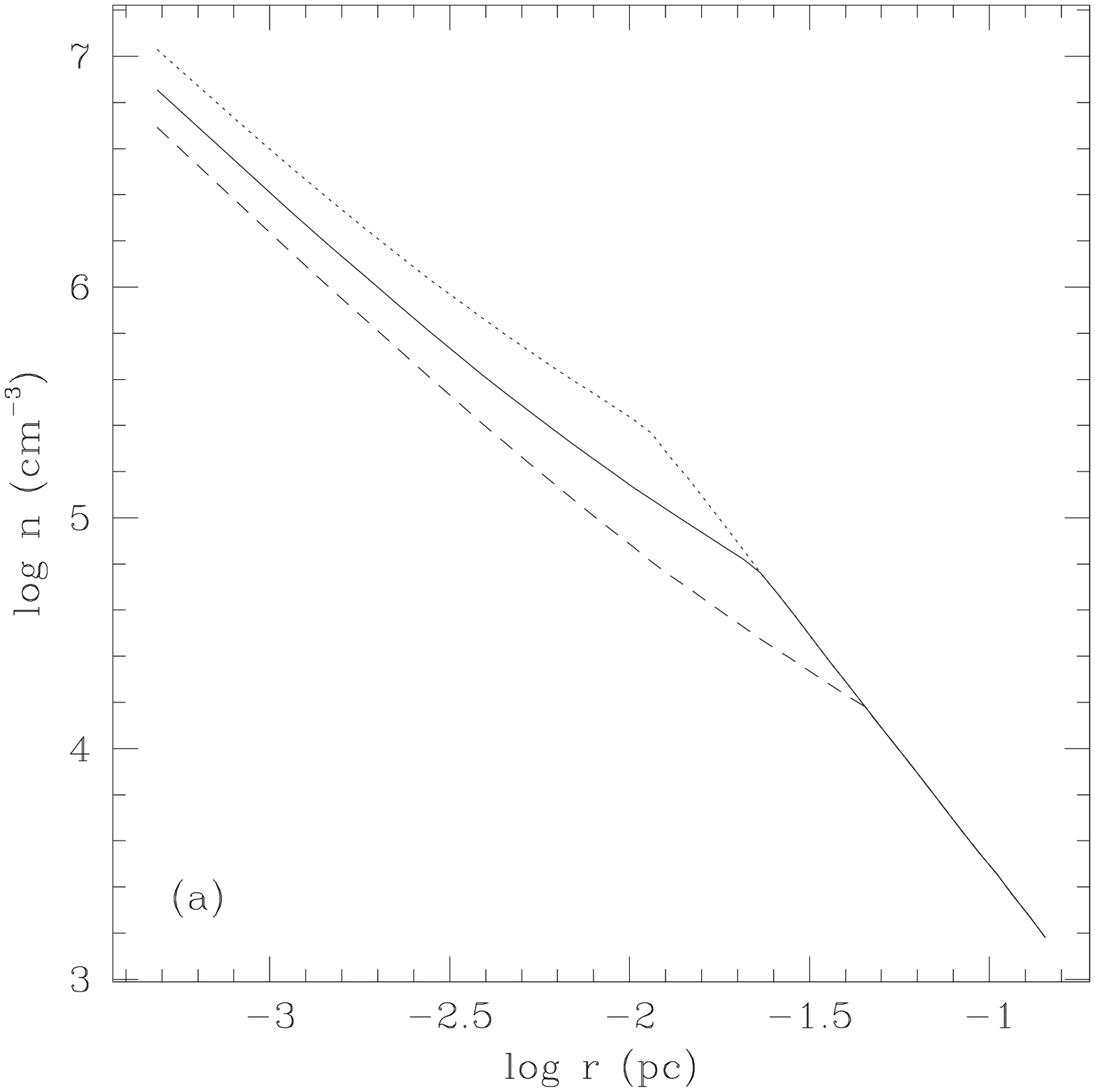}\plotone{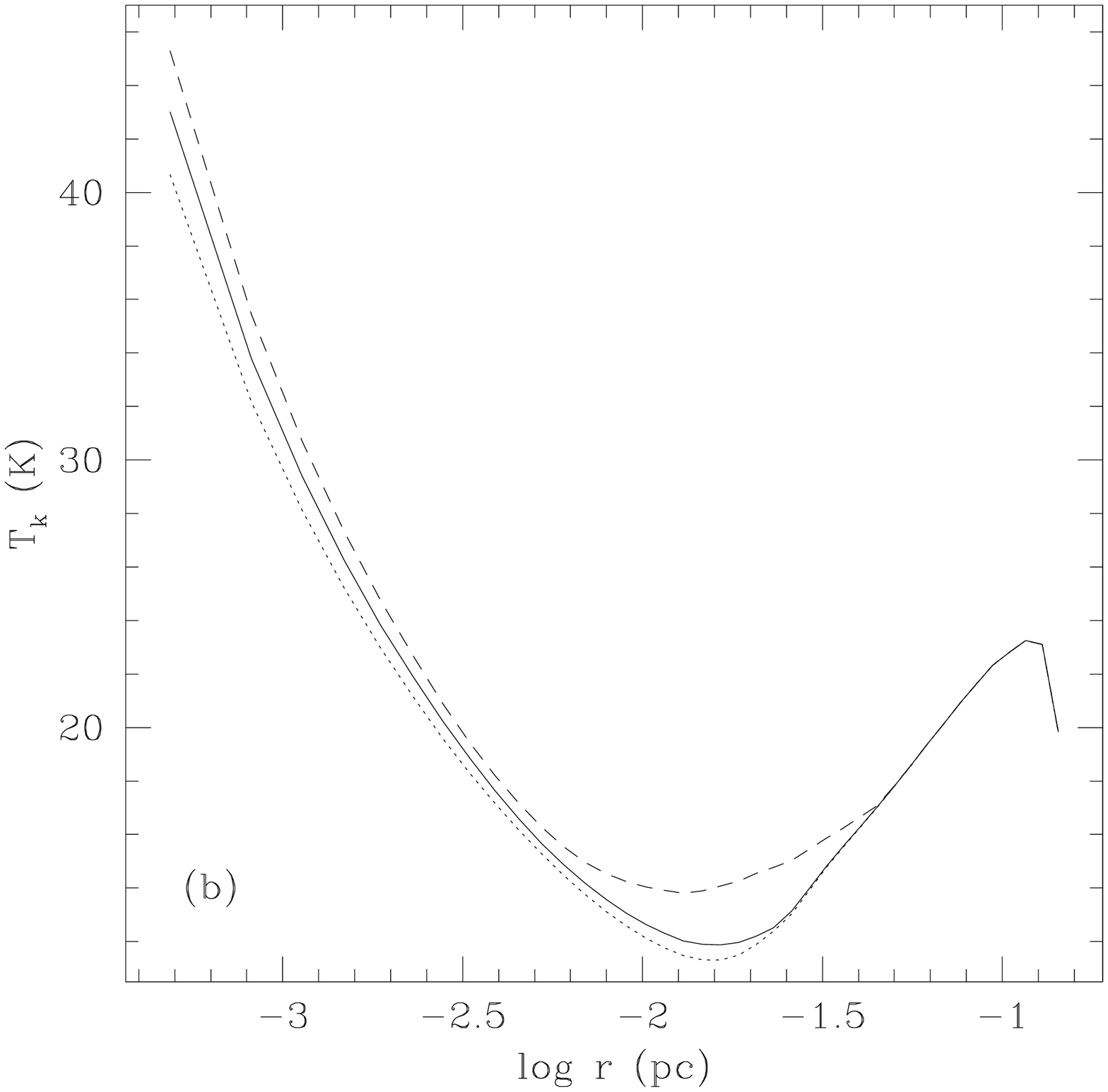}
\plotone{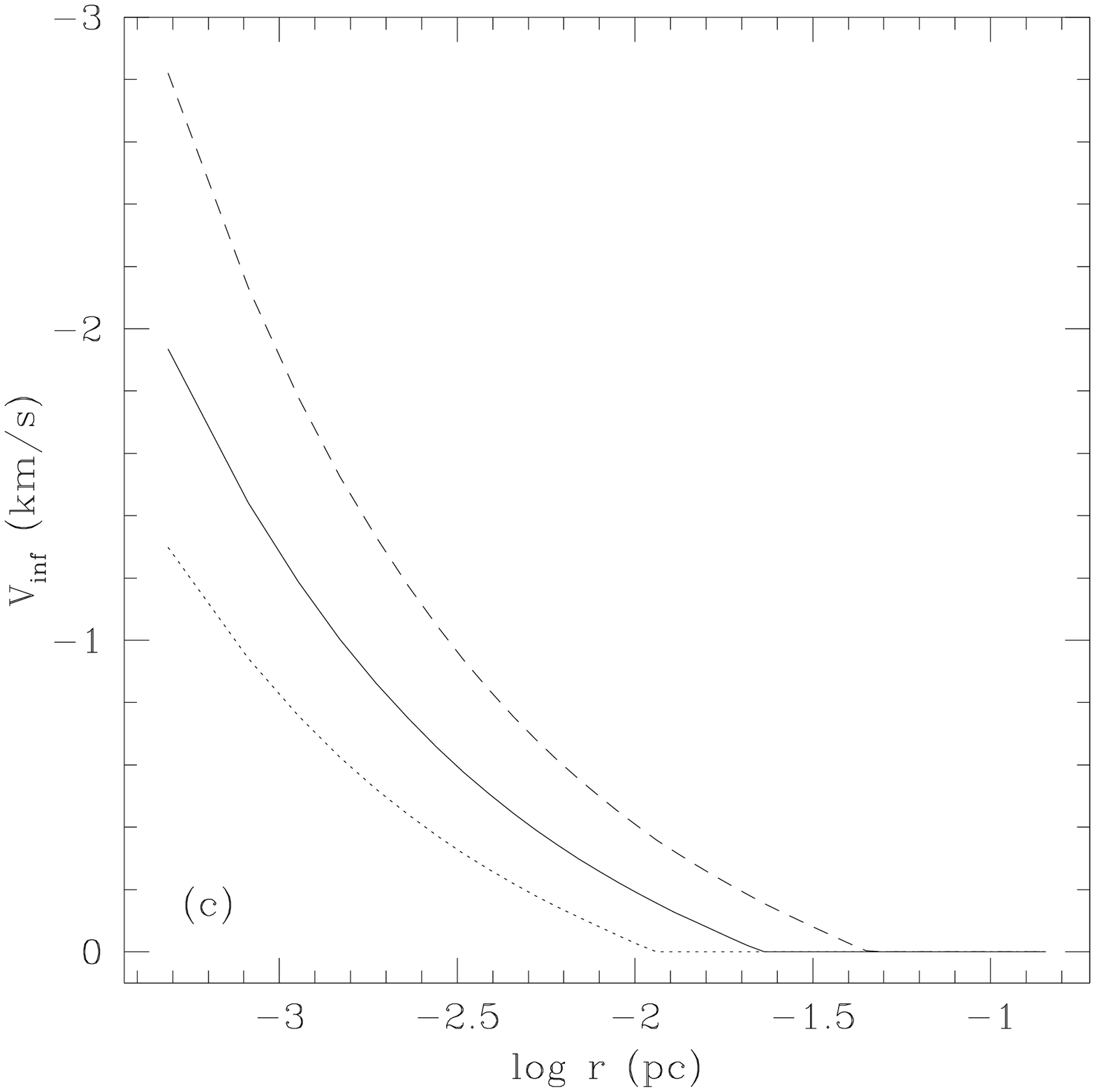}\plotone{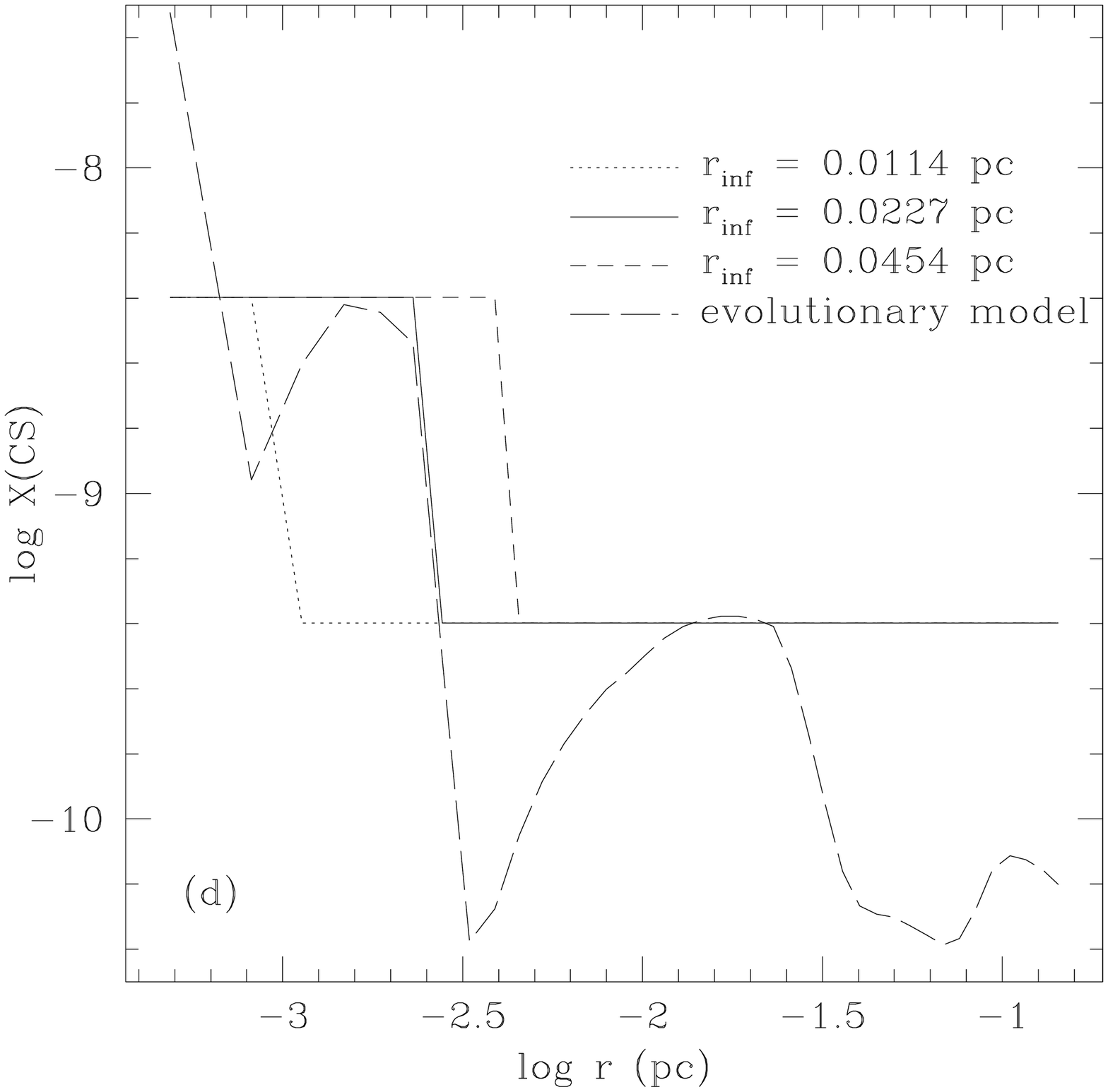}
\epsscale{1.0}
\caption{
Physical properties used for testing different infall radii with the step
models in abundance;
(a) density, (b) kinetic temperature, (c) velocity, and (d) abundance
structure for the three infall radii: 0.0114 (dotted line), 0.0227 (solid),
and 0.0454 (dashed) pc.
In the abundance structures, the long dashed line represents the one from
the evolutionary model, which is considered as a standard model in this test.
}
\end{figure}

\begin{figure}
\figurenum{8}
\epsscale{.6}
\plotone{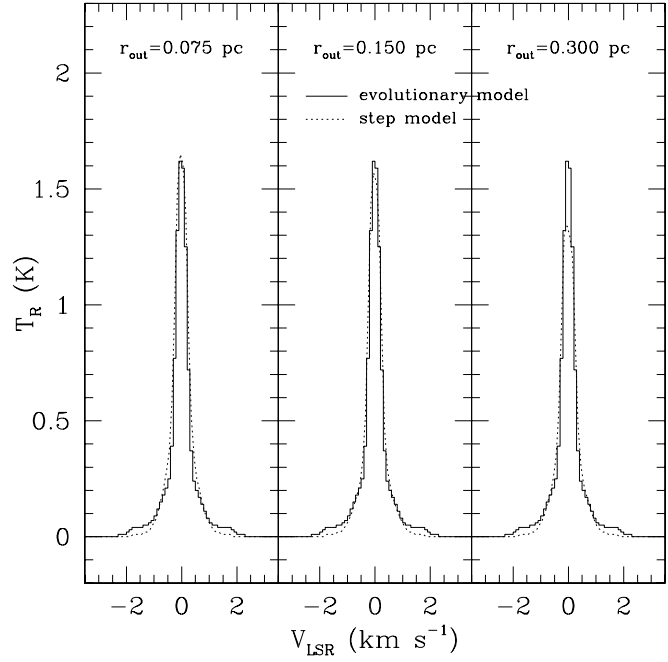}
\caption{
The CS 2$-$1 line profile (solid line) simulated from the evolutionary model
at $t=10^5$ yr is compared with profiles (dotted line) predicted by the step
model with different outer radii but the same infall radius of 0.0227 pc.
The kinetic temperature structure for each model has been calculated
self-consistently.
The beam size for this test is 30$\arcsec$.
The step abundance structure used in this test is the same as the one (dashed
line) in Fig. 3a.
The first panel shows the model with $r_{out}=0.075$ pc; the second panel
shows the model with $r_{out}=0.150$ pc, which is the standard outer radius
for this work; the third panel shows the model with $r_{inf}=0.300$ pc.
The second panel shows the same comparison as the first panel in Fig. 3c
does.
}
\end{figure}

\begin{figure}
\figurenum{9}
\epsscale{.5}
\plotone{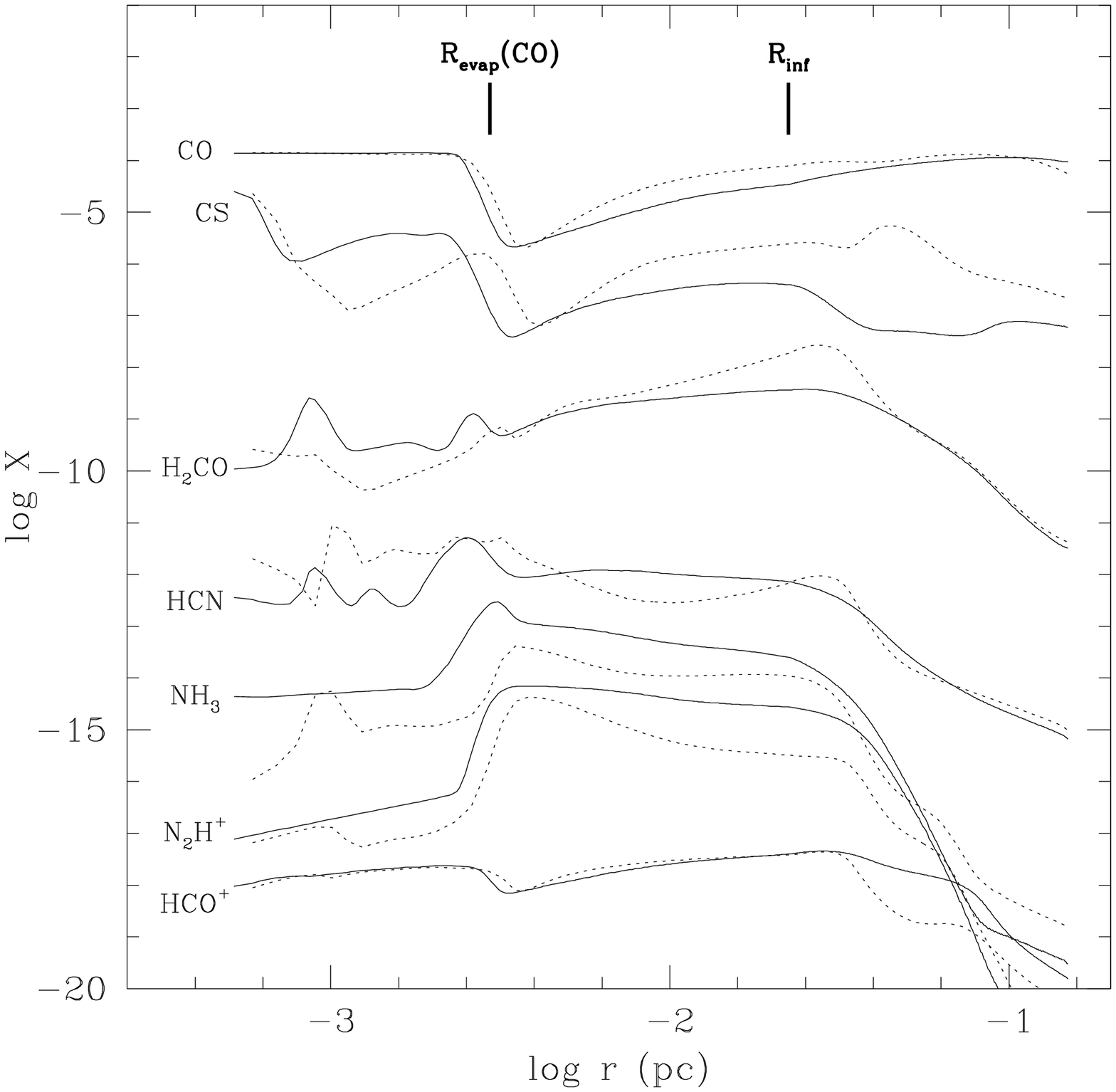}\plotone{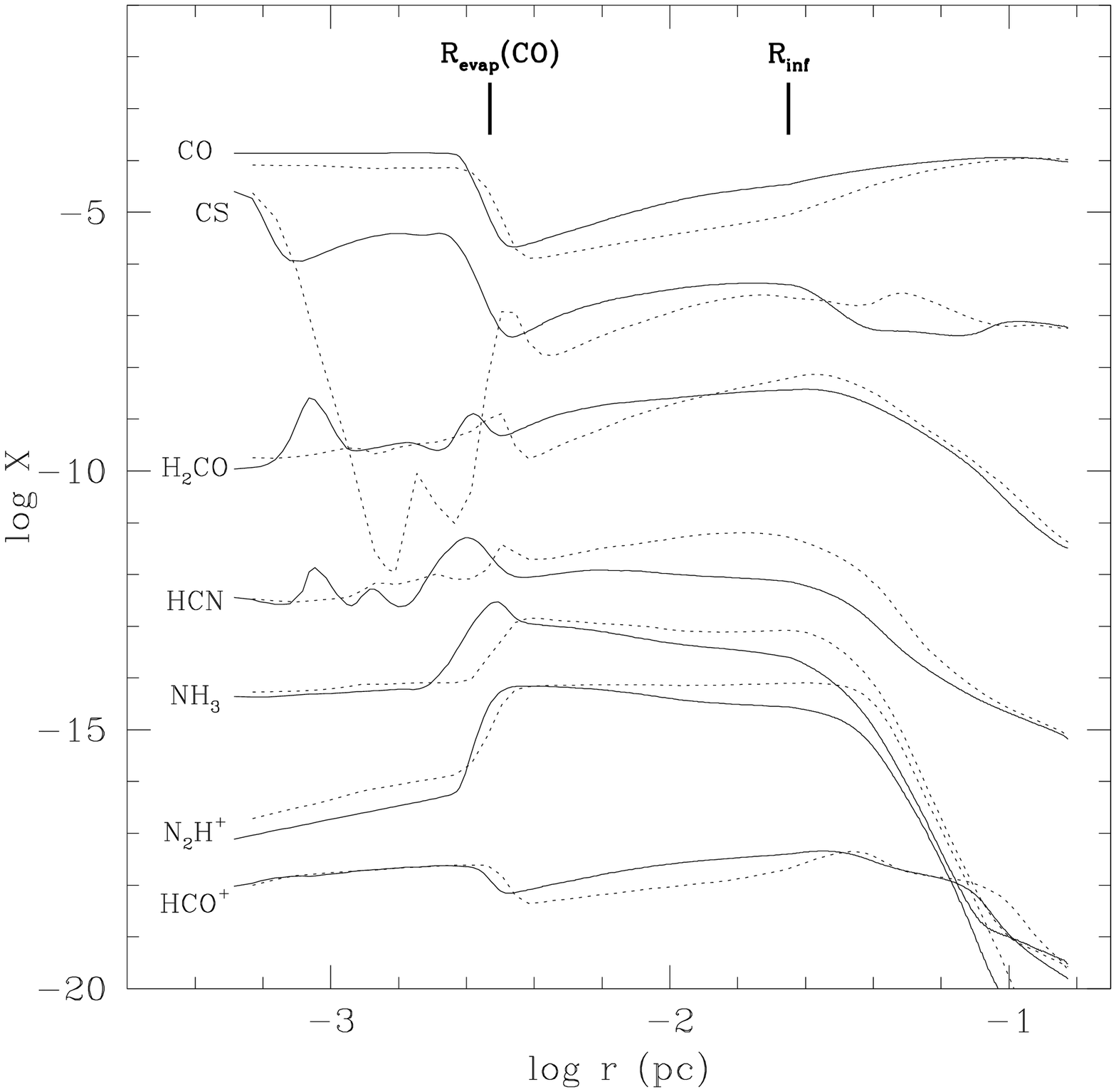}
\epsscale{1.0}
\caption{
The comparisons of abundance profiles calculated by the evolutionary model
(solid lines) at $t=10^5$ yr with the truly static model (dotted line) at
$t=10^5$ yr (left box) and $t=1.1\times 10^6$ yr (right box).
The infall radius and the evaporation front of CO are around 0.023 and
0.003 pc, respectively, and marked with thick vertical lines.
Profiles have been shifted up and down for better comparison.
CS is shifted up by 3.0 orders of magnitude, and H$_2$CO, HCN,  NH$_3$,
N$_2$H$^+$, and HCO$^+$ are shifted down by 0.5, 3.5, 6.0, 5.5, and 9.0 
orders of magnitude, respectively. 
}
\end{figure}

\begin{figure}
\figurenum{10}
\plotone{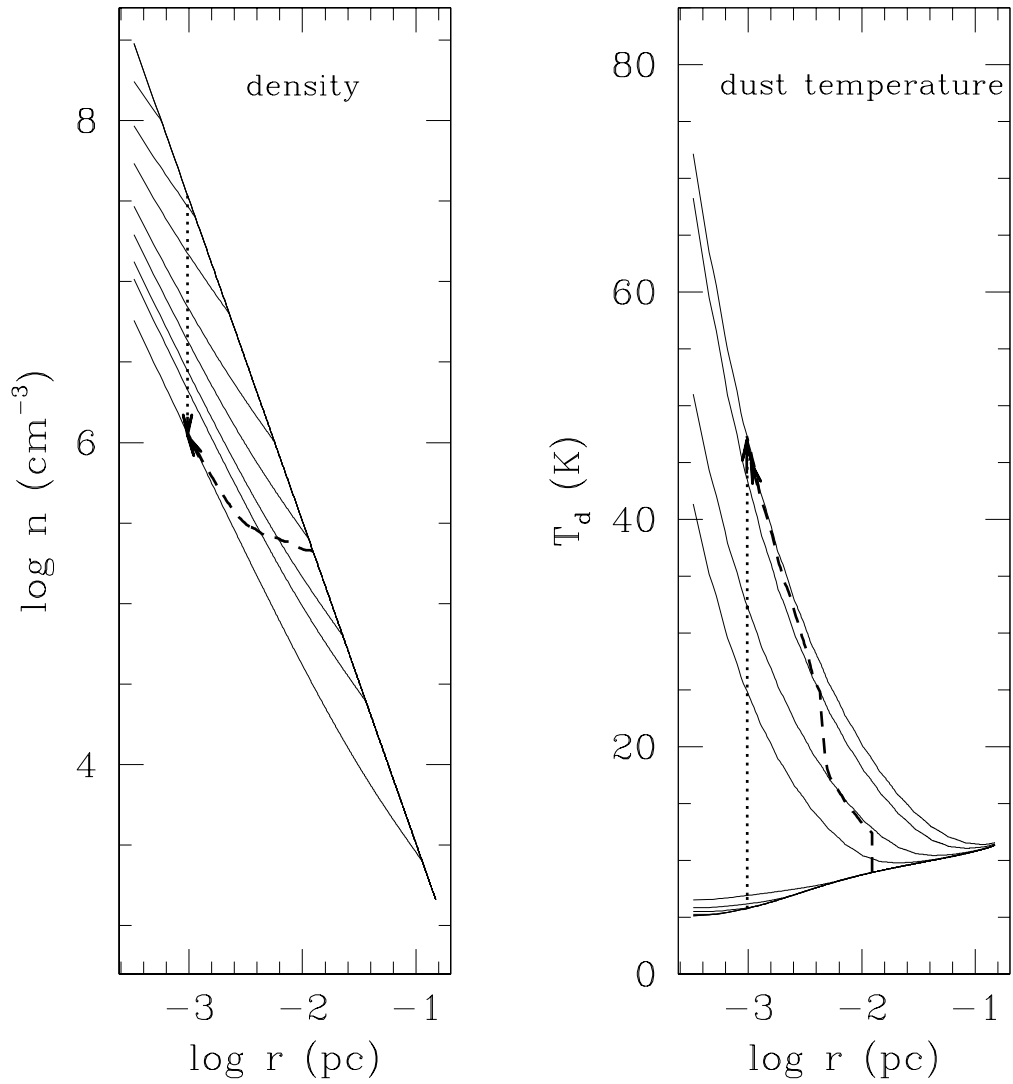}
\caption{
Two different evolutionary histories of density and temperature in the static
model and the evolutionary model. The solid lines represent the density (left
box) and temperature (right box) structures at the time steps of
0, $2.5\times 10^3$, $5\time 10^3$, $10^4$, $2.5\times 10^4$,
$5\times 10^4$, $10^5$, $1.6\times 10^5$, and $5\times 10^5$ yr.
The dotted and dashed lines indicate the routes along which the static and
evolutionary models, respectively, calculate the chemical evolution.
In the static model, the chemical evolution
is calculated at a given radius, so density decreases with time (from top to
bottom), but temperature increases with time (from bottom to top).
However, in the evolutionary model, the chemical evolution
is calculated following a gas parcel, which falls toward the center (from right
to left). Consequently, both density and temperature increase with time.
}
\end{figure}

\begin{figure}
\figurenum{11}
\epsscale{.6}
\plotone{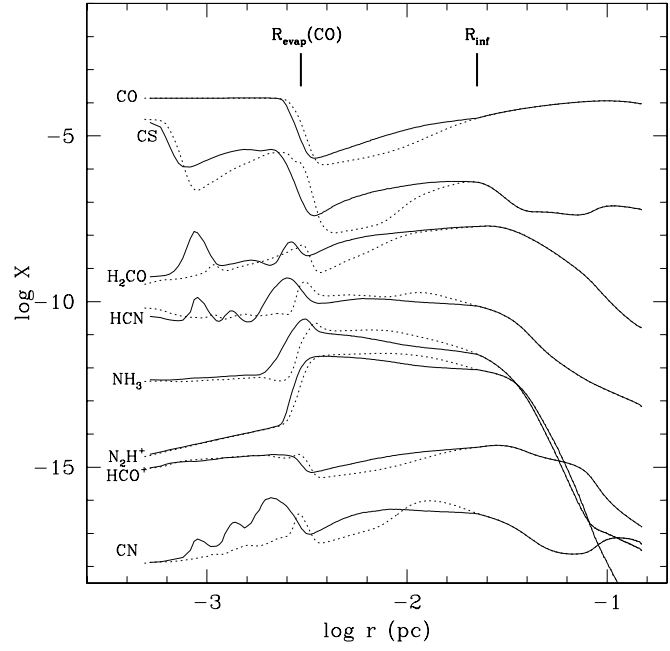}
\caption{
The comparison of abundance profiles calculated by the evolutionary model
(solid lines) and the static model (dotted lines) at $t=10^5$ yr.
The infall radius and the evaporation front of CO are around 0.023 and
0.003 pc, respectively.
Profiles have been shifted up and down for better comparison.
CS, H$_2$CO are shifted up by 3.0 and 0.2 orders of magnitude, and
HCN, NH$_3$, N$_2$H$^+$, HCO$^+$, and CN shifted down by 1.5, 4.0, 3.0, 6.0,
and 8.0 orders of magnitude, respectively.
}
\end{figure}

\begin{figure}
\figurenum{12}
\epsscale{.6}
\plotone{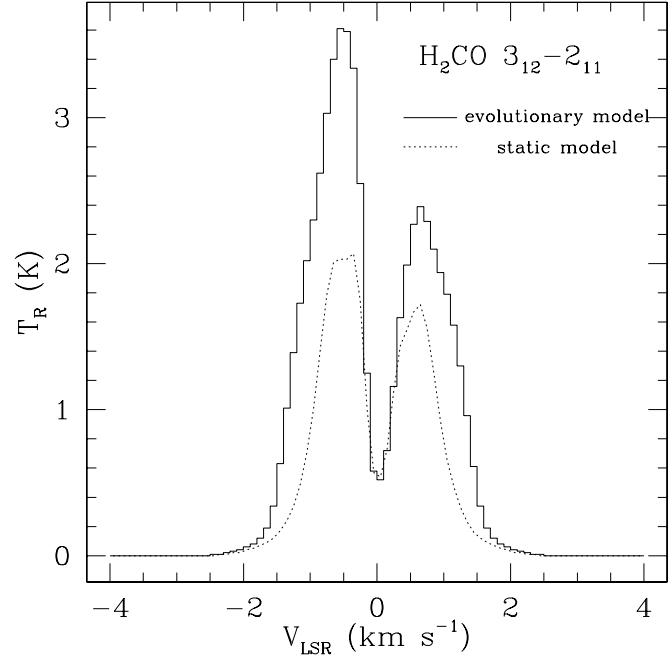}
\caption{
The comparison of the ortho-H$_2$CO line profiles resulting from
the evolutionary model (solid line) and the static model (dotted line)
at $t=10^5$ yr.
The abundance profiles used for the line profiles are plotted in Fig. 11.
The spatial resolution is 3\as.
}

\end{figure}

\end{document}